\documentclass[11pt]{article}
\usepackage{amsmath, amsthm, amsfonts}
\usepackage[margin=1in,footskip=0.25in]{geometry}
\usepackage{graphicx}

\theoremstyle{theorem}
\newtheorem{theorem}{Theorem}
\newtheorem{lemma}[theorem]{Lemma}

\newtheoremstyle{defi}
  {10pt}          
  {10pt}  
  {\rm}  
  {\parindent}     
  {\bf}  
  {. }    
  { }    
  {}     
\theoremstyle{defi}

\begin{document}

\date{}

\title{\bf Strength of the singularities, equation of state and asymptotic expansion}
\author{G. C. Samanta$^{1}$, Mayank Goel$^{2}$ and R. Myrzakulov\\
$^{1, 2}$ Department of Mathematics\\
Birla Institute of Technology and Science (BITS) Pilani,\\
K K Birla Goa Campus,\\
Goa-403726, INDIA,\\ gauranga81@gmail.com\\
$^{3}$Eurasian International Center for Theoretical Physics\\ and Department of General Theoretical Physics,\\ Eurasian National University, Astana 010008, Kazakhstan\\
rmyrzakulov@gmail.com
}

\maketitle

\begin{abstract}In this paper an explicit cosmological model which allows cosmological singularities are discussed. The generalized power-law and asymptotic expansions of the baro-tropic fluid index $\omega$ and equivalently the deceleration parameter $q$, in terms of cosmic time $'t'$ are considered. Finally, the strength of the found
singularities is discussed.
\end{abstract}



\section{Introduction}It is well known that, the singularities are very common problems in general relativity. From the observational data, it is observed that the expansion of our universe is in accelerating way
(Riess \emph{et al.} \cite{1}, Perlmutter \emph{et al.} \cite{2}, Spergel \emph{et al.} \cite{3}, \cite{4}). However, these cosmological puzzlings do not absolutely fit to our current theoretical work.
Therefore, there are two methods of attempt to amend it. One idea is the modifications of general relativity as the correct theory
of gravity (Durrer and Maartens \cite{5}, Nojiri and Odintsov \cite{6}, Starobinsky \cite{7}, Tsujikawa \cite{8}, Nojiri and Odintsov (\cite{9}, \cite{10}), Capozziello \emph{et al.} \cite{11}, Bamba \emph{et al.} \cite{12}). Also, the other major idea assumes the validity of general relativity and postulates the existence of an exotic component in the content of the universe known as dark energy (Padmanabhan \cite{13}, Sahni and Starobinsky \cite{14}).

\par
After the discovery of the expansion of the universe in accelerating way, deeper studies of the phenomenon
 of the dark energy showed the plethora of new singularities (``exotic" singularities) different from big-bang. It is well known that, the cosmological singularities are
 a very interesting problem in general relativity. Hawking and Penrose \cite{15} and Geroch \cite{16} state that, the primary characteristic of a physical singularity is the
 beyond of inextensibility of geodesics. However, the nature of geodesics is not sufficient to capture the detailed features of singularities and
 distinguish physical from un physical ones. Therefore, singularities are classified in terms of strong and weak type (Ellis and Schmidt \cite{17}, Tipler \cite{18}). In a strong singularity, the tidal forces cause complete destruction of objects irrespective of their physical characteristics, whereas a singularity is considered to be weak if the tidal forces are not strong enough to forbid the passage of objects or detectors. In cosmological models, the big-bang singularity is the one example of strong singularity. An example of a weak singularity is the shell crossing singularity in gravitational collapse scenarios where even though curvature invariants diverge, ``strong detectors" can pass the external event (Seifert \cite{19}).
  Apart from this, firstly, a big-rip associated with the phantom dark energy studied by Caldwell \cite{20}. The cosmological models involve with singularities are discussed by Dabrowski \emph{et al.} \cite{21}, and further the classification of singularities are discussed by (Nojiri \emph{et al.} \cite{22}, Bamba \emph{et al.} \cite{23}). Afterward, a sudden
  future singularity or type-II singularity discussed by (Barrow \emph{et al.} \cite{R1}, Barrow \cite{24}, Nojiri and Odintsov \cite{25}, 
  Barrow and Tsagas \cite{R3}, Barrow \emph{et al.} \cite{R4} and Barrowwith and Graham \cite{R5}).  Nevertheless, the singularities which
  fall outside this classification (Kiefer \cite{26}) are curvature singularity with respect to a parallel propagated basis, which show up as directional singularities (Fernandez-Jambrina \cite{27})
  and also intensively studied recently: the little-rip singularities (Frampton \emph{et al.} \cite{28}), and the pseudo-rip singularities (Frampton \emph{et al.} \cite{29}). All the above singularities are
  characterized by violation of all, some or none of the energy conditions which results in a blow-up of all or some of the appropriate physical quantities
  such as: the scale factor, the energy density, the pressure, and the barotropic index (Dabrowski and Denkiewiez \cite{30}). There are three energy conditions: the null
  ($\rho c^2 + p\ge 0$), weak ($\rho c^2 \ge 0$ and $\rho c^2 + p \ge 0$), strong ($\rho c^2 + p\ge 0$ and $\rho c^2+3p\ge0$), and dominant
  energy ($\rho c^2\ge 0$, $-\rho c^2\le p\le\rho c^2$), where $c$ is the speed of light, $\rho$ is the energy density, and $p$ is the pressure.
  \par

  Keeping with the view of the above discussion our work is to look at the classification of singularities involve with the cosmological model in general relativity.

\section{Equations of motion, solutions and singularities}
The metric representation of the Kaluza-Klein space time (Ozel \emph{et al.} \cite{31}) is written as
\begin{eqnarray}
\label{eqn1}
ds^2=dt^2-R^2(t)\left(\frac{dr^2}{1-kr^2}+r^2\left(d\theta^2+\sin^2\theta d\phi^2\right)+
(1-kr^2)d\psi^2\right)
\end{eqnarray}
where $R(t)$ is the scale factor. There are only three distinct possibilities for the geometry, namely $k=-1, 0, 1$ corresponding to the open, flat and closed model of the universe respectively. The source of the gravitational field is assumed to be perfect fluid which is defined by
\begin{equation}
\label{eqn2}
T_{\mu\nu}=(p+\rho)u_{\mu}u_{\nu}-g_{\mu\nu}p,~~~~  (\mu, \nu=0, 1, 2, 3, 4)
\end{equation}
where $u_{\mu}$ is the five velocity vector, satisfying $u_{\mu}u^{\mu}=1$.
The Einstein field equations can be written as
\begin{equation}
\label{eqn3}
R_{\mu\nu}-\frac{1}{2}g_{\mu\nu}R=T_{\mu\nu}.
\end{equation}
Here, the units to be considered as $c=1=8\pi G$.
Using equations \eqref{eqn1} and \eqref{eqn2} in \eqref{eqn3}, it follows that
\begin{equation}
\label{eqn4}
6\left(\frac{\dot{R}}{R}\right)^2+6\frac{k}{R^2}=\rho,
\end{equation}
\begin{equation}
\label{eqn5}
-3\frac{\ddot{R}}{R}-3\left(\frac{\dot{R}}{R}\right)^2-3\frac{k}{R^2}=p.
\end{equation}
For the flat model ($k=0$), we have
\begin{equation}
\label{eqn6}
6\left(\frac{\dot{R}}{R}\right)^2=\rho,
\end{equation}
\begin{equation}
\label{eqn7}
-3\frac{\ddot{R}}{R}-3\left(\frac{\dot{R}}{R}\right)^2=p.
\end{equation}
The overhead dot stands for ordinary derivative with respect to time co-ordinate.
Dividing \eqref{eqn7} by \eqref{eqn6}, we get
\begin{equation}
\label{eqn8}
\frac{p}{\rho}=-\frac{R\ddot{R}}{2{\dot{R}^2}}-\frac{1}{2}.
\end{equation}
Now, the time dependent baro-tropic fluid index $\omega (t)$ can be defined as the ratio of the pressure and the energy density of the universe and which can be written as
\begin{equation}
\label{eqn9}
\omega (t)=\frac{p}{\rho}= -\frac{R\ddot{R}}{2{\dot{R}^2}}-\frac{1}{2}.
\end{equation}
Let us now define the deceleration parameter
\begin{equation}
\label{eqn10}
q=-\frac{R{\ddot{R}}}{\dot{R}^2}.
\end{equation}
From \eqref{eqn9} and \eqref{eqn10}, we have
\begin{equation}
\label{eqn11}
q=2\omega +1,
\end{equation}
\begin{equation}
\label{eqn12}
\omega=\frac{q-1}{2}.
\end{equation}
Let us define a non-linear time dependent function $f=\ln(R)$
\begin{equation}
\label{eqn13}
\frac{\ddot{f}}{\dot{f}^2}=-2(\omega +1)=-(q+1).
\end{equation}
We define,
\begin{equation}
\label{eqn14}
g(t)=-2(1+\omega (t))=-(q(t)+1).
\end{equation}
By the help of equation \eqref{eqn14}, the equations \eqref{eqn11} and \eqref{eqn12} reduce to
\begin{equation}
\label{eqn15}
\omega (t)=-\frac{g(t)}{2}-1,
\end{equation}
\begin{equation}
\label{eqn16}
q(t)=-g(t)-1.
\end{equation}
From \eqref{eqn13} and \eqref{eqn14}, we can have
\begin{equation}
\label{eqn17}
{\left(\frac{1}{\dot{f}}\right)}^{\cdot}=g(t).
\end{equation}
Integrating \eqref{eqn17}, we get
\begin{equation}
\label{eqn18}
\dot{f}=\left(\int g(t)dt+k_1\right)^{-1},
\end{equation}
which can be solved with two free constants $k_1$ and $k_2$,
\begin{equation}
\label{eqn19}
R(t)=\exp\left(\int\left(\int g(t)dt+k_1\right)^{-1}dt+k_2\right).
\end{equation}
The constant $k_2$ is the part of a global constant factor $R(t_0)=\exp(k_2)$,
\begin{equation}
\label{eqn20}
R(t)=R(t_0)\exp\left(\int_{t_0}^{t}\left(\int g(t)dt+k_1\right)^{-1}dt\right),
\end{equation}
models with this type of exponential behavior can be found in (Dabrowski and Marosek \cite{32}).
Performing the Friedman equations \eqref{eqn6} and \eqref{eqn7},
\begin{equation}
\label{eqn21}
\rho(t)=6\left(\int_{t_0}^{t}g(t)dt+k_1\right)^{-2},
\end{equation}
\begin{equation}
\label{eqn22}
p(t)=\frac{3(g(t)-2)}{\left(\int_{t_0}^{t}g(t)dt+k_1\right)^2},
\end{equation}
where $k_1=\sqrt{6}\rho(t_0)^{-\frac{1}{2}}$, if $\rho$ is infinity at $t=t_{0}$, then in this case $k_1=0$.
Hence, expression of the scale factor reduces to
\begin{equation}
\label{eqn23}
R(t)=\exp\left(\int\frac{dt}{\int g(t)dt}\right),
\end{equation}
The rate of growth of the function $g(t)$ has several qualitative behaviors. Let us assume the function $g(t)$ has a power series expansion around the point $t=0$,
\begin{equation}
\label{eqn24}
g(t)=g_{0}t^{n_0}+g_1t^{n_1}+g_2t^{n_2}+\cdots\cdots,~~~~~   n_0<n_1<\cdots
\end{equation}
The scale factor, energy density and the pressure are obtained as:
\begin{equation}
\label{eqn25}
f(t)=  \begin{cases} -\frac{n_{0}+1}{g_{0} n_{0}}t^{-n_{0}}-\frac{(n_{0}+1)^2 g_{1}}{(n_{1}+1)(n_{1}-2n_{0})g_{0}^2}t^{n_{1}-2n_{0}} &\mbox{if  } n_0\neq -1, 0  \\
-\frac{t}{g_{0}}-\frac{(g_{0}+g_{1})}{2g_{0}^2}t^2 & \mbox{if  }  n_0=-1,~~ |t|\le 2 \\
\frac{ln t}{g_0}-\frac{g_1}{2(g_0)^2}t & \mbox{if  } n_0=0,
\end{cases}
\end{equation}
\begin{equation}
\label{eqn26}
R(t)=\begin{cases}\exp\left(-\frac{(n_0+1)}{g_0n_0}t^{-n_0}
-\frac{g_1(n_0+1)^2}{g_0^2(n_1+1)(n_1-2n_0)}t^{n_1-2n_0}\right) & \mbox{if  } n_0\neq -1, 0\\
\exp\left(-\frac{t}{g_0}-\frac{(g_0+g_1)}{2g_0^2}t^2\right) & \mbox{if  } n_0=-1, |t|\leq2 \\
\exp\left(\frac{lnt}{g_0}-\frac{g_1}{2g_0^2}t\right)& \mbox{if  }  n_0=0
\end{cases}
\end{equation}
\begin{equation}
\label{eqn27}
\rho(t)=\begin{cases}6\frac{(n_0+1)^2}{g_0^2}t^{-2n_0-2}+6\frac{g_1^2(n_0+1)^4}{g_0^4(n_1+1)^2}t^{2n_1-4n_0-2}
+12\frac{(n_0+1)^3g_1}{g_0^3(n_1+1)}t^{n_1-3n_0-2} & \mbox{if  } n_0\neq -1, 0\\
6\frac{g_0^2g_1^2}{g_0^4}t^2+12\frac{g_0g_1}{g_0^3}t+\frac{6}{g_0^2} & \mbox{if  } n_0=-1\\
\frac{6}{g_0t^2}-12\frac{g_1}{2g_0^3t}+3\frac{g_1^2}{g_0^4} & \mbox{if  } n_0=0
\end{cases}
\end{equation}
and
\begin{equation}
\label{eqn28}
p(t)=\begin{cases}
3\frac{(n_0+1)^2}{g_0}t^{-n_0-2}+3\frac{g_1(n_0+1)^2(n_1-2n_0-1)}{g_0^2(n_1+1)}t^{n_1-2n_0-2}
-6\frac{(n_0+1)^2}{g_0^2}t^{2n_0-2} &  \\
-6\frac{g_1^2(n_0+1)^4}{g_0^4(n_1+1)^2}t^{-2n_1-4n_0-2}
+12\frac{g_1(n_0+1)^3}{g_0^3(n_1+1)}t^{n_1-3n_0-2} & \mbox{if  } n_0\neq -1, 0\\
\frac{3g_0g_1-6}{g_0^2}-\frac{12g_1}{g_0^2}t-\frac{6g_1^2}{g_0^2}t^2 & \mbox{if  } n_0=-1\\
\frac{6g_1}{g_0^3t}-\frac{3g_0-6}{g_0^2t^2}-\frac{6g_1^2}{4g_0^2} & \mbox{if  } n_0=0.
\end{cases}
\end{equation}
Please see the behavior of the pressure and energy density with time from the figure 1, 2 \& 3 which describes the different phases of the universe. We consider the following five possibilities for the parameter $n_0$. These singularities are classified in following way (Nojiri et al. \cite{22}, Dabrowski and Denkiewiez \cite{33}, Fernandez-Jambrina \cite{34}):
\begin{itemize}
\item For $n_0<-2$, both pressure $(p)$ and energy density $(\rho)$ vanish at $t=0$, the scale factor $R(t)$ becomes constant, whereas the baro-tropic fluid index $\omega$ diverges.
These are called type-IV singularities.
\item For $n_0=-2$, the energy density $(\rho)$ vanishes at $t=0$, the pressure $(p)$ and the scale factor $(R)$ becomes finite, whereas the baro-tropic fluid index $\omega$ diverges, which is called special case of generalized sudden singularities. 

\item For $n_0\in (-2, -1]$, the energy density $(\rho)$, the pressure $(p)$, the scale factor $(R)$ and $\omega$ all are finite for $t=0$.
Hence, there is no singularities within this range.
\item For $n_0\in (-1, 0]$, the energy density $(\rho)$, the pressure $(p)$ and $\omega$ diverge at $t=0$, whereas the scale factor $(R)$ vanishes.
These are called type-III, Big Freeze of finite scale factor singularities.

\item For $n_0>0$, the energy density $(\rho)$ and the pressure $(p)$ diverges at $t=0$ as $t^{-2(n_0+1)}$ and the baro-tropic fluid index 
$\omega\rightarrow -1$. 
We may name these, grand rip or grand bang/crunch, depending on the behavior of
the scale factor at the singular point.
\end{itemize}
\section{Behavior of the model at infinite time}
Apart from the above discussion of the singularities at a finite time $'t'$, we may analyze the behavior of the model at $t \to\infty$. For this
observations we can think about the asymptotic behavior of $g(t)$ for large $t$. If, we take $t_0\to\infty$, then equations \eqref{eqn20}, \eqref{eqn21} and \eqref{eqn22} reduce to

\begin{equation}
\label{eqn29}
R(t)=\exp\left(-\int\left(\int_{t}^{\infty}g(t)dt+k_1\right)^{-1}dt\right),
\end{equation}
\begin{equation}
\label{eqn30}
\rho (t)=6\left(\int_{t}^{\infty}g(t)dt+k_1\right)^2
\end{equation}
and
\begin{equation}
\label{eqn31}
p(t)=\frac{3(g(t)-2)}{\left(\int_{t}^{\infty}g(t)dt+k_1\right)^2}.
\end{equation}
If the constant $k_1=0$, then the energy density $(\rho)$ and the pressure $(p)$ diverges at $t\to\infty$. The above expressions for scale factor $(R)$, energy density $(\rho)$ and pressure $(p)$ are well define if the integral
\begin{equation}
\label{eqn32}
\int_{t}^{\infty}g(t)dt
\end{equation}
is convergent. This ensure that $k_1=\sqrt6 (\rho(\infty))^{-\frac{1}{2}}$, which is useful for controlling the asymptotic nature of $\rho$ and $p$.
\begin{lemma}
A necessary and sufficient condition for the convergence of the integral \eqref{eqn32}, i. e. $\int_{t_1}^{\infty}g(t)dt$, where $g$ is positive in $[t_1, t]$
is that there exists a positive number $K$ independent of $t$, such that $\int_{t_1}^{t}g(t)dt\leq K$, for any $t\ge t_1$. The integral
$\int_{t_1}^{\infty}g(t)dt$ is said to be convergent if $\int_{t_1}^{t}g(t)dt$ tends to constant as $t\rightarrow\infty$.
\end{lemma}
\textbf{Proof:} Since $g$ is positive in $[t_1, t]$ the positive function of $t$, $\int_{t_1}^{t}gdt$ is monotone increasing as $t$ increases
and will therefore tend to a finite limit if and only if it is bounded above. That is, there exist a positive number $K$, independent of $t$,
 such that $\int_{t_1}^{t}g(t)dt\leq K$, for every $t\ge t_1$. \par
 If no such type of number $K$ exist, the monotonic increasing function $\int_{t_1}^{t}g(t)dt$ is not bounded above and therefore tends to $\infty$,
 as $t\rightarrow\infty$ and so $\int_{t_1}^{t}g(t)dt$ diverges to $\infty$.\hfill $\Box$
 \par
 From the above Lemma, we conclude that the integral $\int_{t}^{\infty}g(t)dt$ is finite, if $g(t)$ is bounded above by $\frac{1}{t}$. This implies
 that $g(t)\rightarrow 0$ for large values of time $'t'$, then from the equation \eqref{eqn15}, we can say that the values of the baro-tropic fluid index $\omega$ is $-1$. Also, based upon above analysis we can conclude the following points:
 \begin{itemize}
 \item If the integral $\int_{t}^{\infty}g(t)dt$ is convergent and the value is positive, then we observe that the scale factor $(R)$ from the \eqref{eqn26} decreases exponentially to zero
 as $t\to\infty$. It would be a sort of little crunch. $\omega_{\infty}=-1$ is the asymptotic value of the baro-tropic fluid index $\omega$. 
 At infinity, $R(t)$ is an integrable function, hence this case is included in the set of directional singularities described by (Fernandez-Jambrina \cite{27}), which are called strong singularities, but only easy to reach
 for some observers.
\item If the integral $\int_{t}^{\infty}g(t)dt$ is convergent and the value of the integral is negative, then the scale factor $(R)$ grows up exponentially at infinity. It is the Little Rip (Frampton \emph{et al.} \cite{28}, \cite{37}), or for different types of $g(t)$, the Little Sibling (Bouhmadi-Lopez \emph{et al.} \cite{38}).
\item For $k_1\neq 0$, the physical parameters $R$, $\rho$ and $p$ are obtained from the equations \eqref{eqn20}, \eqref{eqn21} and \eqref{eqn22} are well behaved, provided the integral $\int_{t}^{\infty}g(t)dt$ is infinite. In this case both $(\rho)$ and $(p)$ are tend to zero as $t\to \infty$.
    The asymptotic value of the baro-tropic fluid index $\omega_{\infty}$ is $-1$ if $g(t)\to 0$.
    \end{itemize}
\par
Now, we may look for the behavior of causal geodesics discussed by (Hawking and Ellis \cite{39}). Consider the parameterized curves as $\gamma (\tau)=(t(\tau), r(\tau), \theta (\tau), \phi (\tau), \psi(\tau))$, and impose a normalization condition on the velocity $u(\tau)=\gamma'(\tau)$, depending on its causal type
\begin{equation}\nonumber
-R^2(t(\tau))\left(r'^2(\tau)+r^2(\tau)(\phi'^2(\tau)+\sin^2\theta(\tau)\phi'^2(\tau))+\psi'^2(\tau)\right)
+t'^2(\tau)=\varepsilon,
\end{equation}
\begin{equation}
\label{eqn33}
\varepsilon=||\gamma'(\tau)||^2
=\begin{cases} \mbox{Timelike}: -1 \\
\mbox {Lightlike}: 0 \\
\mbox {Spacelike}: +1,
\end{cases}
\end{equation}
where the overhead dash denotes derivative with respect to the parameter $\tau$. 
 \begin{equation}
 \label{eqn34}
 p=u\cdot\partial_{r}=R^2(t)r'.
 \end{equation}
Equation \eqref{eqn34} together with equation \eqref{eqn33} permit to make the system of first order differential equations as follows
\begin{equation}
\label{eqn35}
r'=\frac{p}{R^2(t)},  t'=\sqrt{-\varepsilon+\frac{p^2}{R^2(t)}}
\end{equation}
for the normal parameter $\tau$.\\
We analyze to know whether the causal geodesics are complete (Hawking and Ellis \cite{39}), that is, if the parameter $\tau$ can be extended from $-\infty$ to $\infty$. \\
Here, we restrict our discussion to light-like geodesics only:
\begin{itemize}
\item \textbf {Light-like geodesics}
\end{itemize}
Since in this case $\varepsilon=0$, from \eqref{eqn35}, we have
\begin{equation}
\label{eqn36}
\tau=\frac{1}{p}\int_{0}^{t}R(t)dt.
\end{equation}
Here, $\displaystyle{R(t)=e^{-\frac{g_0\alpha}{t^{n_0}}}}$, the integral is convergent for positive value of $g_0$. This implies that, the light-like geodesics meet the
singularity at $t=0$ in a finite normal time $\tau$. Therefore, these geodesics are incomplete. The integral is not convergent for $g_0\leq 0$, 
and it takes an infinite normal time $\tau$ to reach $t=0$. 
Therefore, this case yields the light-like geodesics avoid reaching the singularity and are complete in that direction.
This is similar to Big-Rip singularities (Fernandez-Jambrina and Lozkoz \cite{36}).
\section{Strength of the singularities}
Ellis and Schmidt \cite{40} introduced the idea of strong singularity. When tidal forces influence a several disruption is called a strong singularity. As per the (Tipler's \cite{41}) idea, when volume tends to zero on approaching the singularity along the geodesics is called a strong singularity.
Whereas the definition of (Krolak \cite{42}) is less restrictive, it is just demands that the derivative of the volume with respect to proper time to be negative. Hence,
there are singularities which are strong according to Krolak's definition, but are weak according to Tipler's. Therefore, this definition has been further
revised by (Rudnicki \emph{et al.} \cite{43}). From these definition it clear that, $R_{\mu\nu}u^{\mu}u^{\nu}$ is non-negative when an observer moving with velocity $u$ for time-like and light-like events. 
\begin{itemize}
\item \textbf{Light-like geodesics:}
\end{itemize}
According to (Clarke and Krolak \cite{44}) a light-like geodesic meets a strong singularity, according to the judgement of Tipler, the singularity is strong at proper time $\tau_0$ if and only if the integral
of the Ricci tensor \\
\begin{equation}
\label{eqn37}
\int_0^{\tau}d\tau'\int_0^{\tau'}d\tau''R_{\mu\nu}u^{\mu}u^{\nu}
\end{equation}
diverges as $\tau$ tends to $\tau_0$. \\
According to Krolak's criterion, a strong singularity meet by light-like geodesic at proper time $\tau_0$ if and only if
the integral
\begin{equation}
\label{eqn38}
\int_0^{\tau}d\tau'R_{\mu\nu}u^{\mu}u^{\nu}
\end{equation}
diverges as $\tau$ tends to $\tau_0$. The velocity of geodesic is defined as $u=(t', r', \theta', \phi', \psi')=\left(\frac{p}{R}, \frac{p}{R^2}, 0, 0, 0\right)$, integral of
\begin{equation}
\label{eqn39}
R_{\mu\nu}u^{\mu}u^{\nu}d\tau=3p^2\left(\frac{R'^2}{R^4}-\frac{R''}{R^3}\right)\frac{Rdt}{p}\simeq
\frac{3pg_0\alpha n_0(n_0+1)}{t^{n_0+2}}e^{\frac{\alpha g_0}{t^{n_0}}}dt
\end{equation}
blows up at $t=0$ for all $g_0>0$ and hence these singularities are strong according to both definitions. For $g_0<0$ we already know that
these geodesics do not even reach the singularity.
\begin{itemize}
\item \textbf{Time-like geodesics:}
\end{itemize}
 As per the usual definitions, it is worthwhile to know that, the singularities encountered by time-like geodesics are strong or not. According
to Tipler's definition, a time-like geodesics meets a strong singularity, at proper time $\tau_0$ if the integral of the Ricci tensor
\begin{equation}
\label{eqn40}
\int_0^{\tau}d\tau'\int_0^{\tau'}d\tau''R{\mu\nu}u^{\mu}u^{\nu}
\end{equation}
blows up as $\tau$ tends to $\tau_0$.\\
Following Krolak's definition, a time-like geodesic meets a strong singularity at proper time $\tau_0$ if the integral
\begin{equation}
\label{eqn41}
\int_0^{\tau}d\tau'R_{\mu\nu}u^{\mu}u^{\nu}
\end{equation}
blows up on approaching to singularity.\\
For co-moving geodesics, $u=(1, 0, 0, 0, 0)$, integrals of
\begin{equation}
\label{eqn42}
R_{\mu\nu}u^{\mu}u^{\nu}d\tau=-\frac{6R''}{R}dt\simeq-\frac{6\alpha^2 n_0^2}{t^{2n_0+2}}dt
\end{equation}
blow up for all $n_0\geq -2$ and hence singularities are strong at $t=0$.\\
For radial geodesics, $u=\left(\sqrt{1+\frac{p^2}{R^2}}, \pm\frac{p}{R^2}, 0, 0, 0\right)$, the analysis is similar.
\begin{equation}
\label{eqn43}
R_{\mu\nu}u^{\mu}u^{\nu}d\tau=\frac{\frac{-6R''}{R}+3p^2\left(\frac{R'^2}{R^4}-\frac{R''}{R^3}\right)}{\sqrt{1+\frac{p^2}{R^2}}}dt \\
\simeq\begin{cases}\frac{-6R''}{p}+3p\left(\frac{R'^2}{R^3}-\frac{R''}{R^2}\right) & \mbox{if  } R\rightarrow 0 \\
\frac{-6R''}{R}+3p^2\left(\frac{R'^2}{R^4}-\frac{R''}{R^3}\right)& \mbox{if  } R\rightarrow \infty
\end{cases}
\end{equation}
For $g_0>0$; $R$, $R''$ tend to zero as $t\to 0$, but the term $p$ is exponentially divergent. \\
The $p$ term approaches to zero and the integrals of the $\frac{R''}{R}$ term is divergent for $g_0<0$. Therefore, radial
geodesics meet a strong singularity in both the cases as $t\to 0$. For $g_0<0$, singularities are strong for all geodesics except for light-like case, which are not even incomplete.
\section{Summary}
Overall, in this paper authors proposed the present behavior of the universe and classify some singularity by the help of generalized power and asymptotic expansions of the baro-tropic fluid equation of state of index $\omega$ and the deceleration parameter $q$ in terms of cosmic time 't'.  We classified the types of singularities into four classes for finite time. The generalized sudden or type-IV singularities are obtained for for $n_0<-2$. The special case of generalized sudden singularities are obtained for $n_0=-2$, . For $n_0\in (-2, -1]$, there is no singularities within this range. The type-III, Big Freeze of finite scale factor singularities are obtained for $n_0\in (-1, 0]$. The grand rip or grand bang/crunch singularity (it depends on the behavior of the scale factor at the singular point)  For $n_0> 0$ is obtained. Finally, we concluded our result with the strength of the singularities and for all geodesics singularities are strong except light-like geodesics.

\begin{figure}[ht!]
  \centering
  \includegraphics[width=11cm,height=8cm]{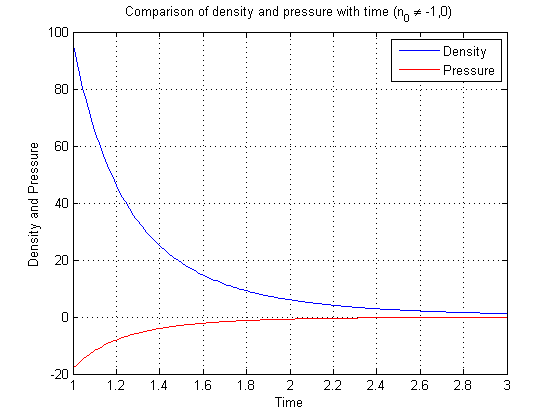}
 \caption{This figure indicates the variation of energy density and pressure with time. From the figure it is observed that the rapid expansion
 of the universe has occurred in first phase i. e. in initial epoch which is known as inflationary period of the universe. The present epoch is described by an accelerated expansion phase because of negative pressure.}
\end{figure}

\begin{figure}[ht!]
  \centering
  \includegraphics[width=11cm,height=8cm]{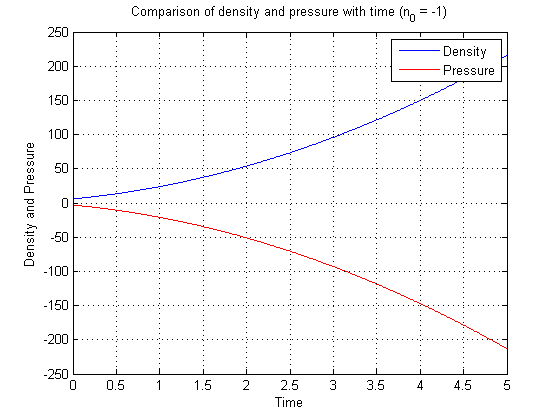}
 \caption{This figure indicates that the variation of energy density and pressure with time. From this figure it seems that, the universe always characterized by an accelerated expansion i. e. the rate of expansion of the universe increases for ever because the pressure always negative and tends to negative infinity for infinite time. Hence, in this case, there is no deceleration phase. }
\end{figure}

\begin{figure}[ht!]
  \centering
  \includegraphics[width=11cm,height=8cm]{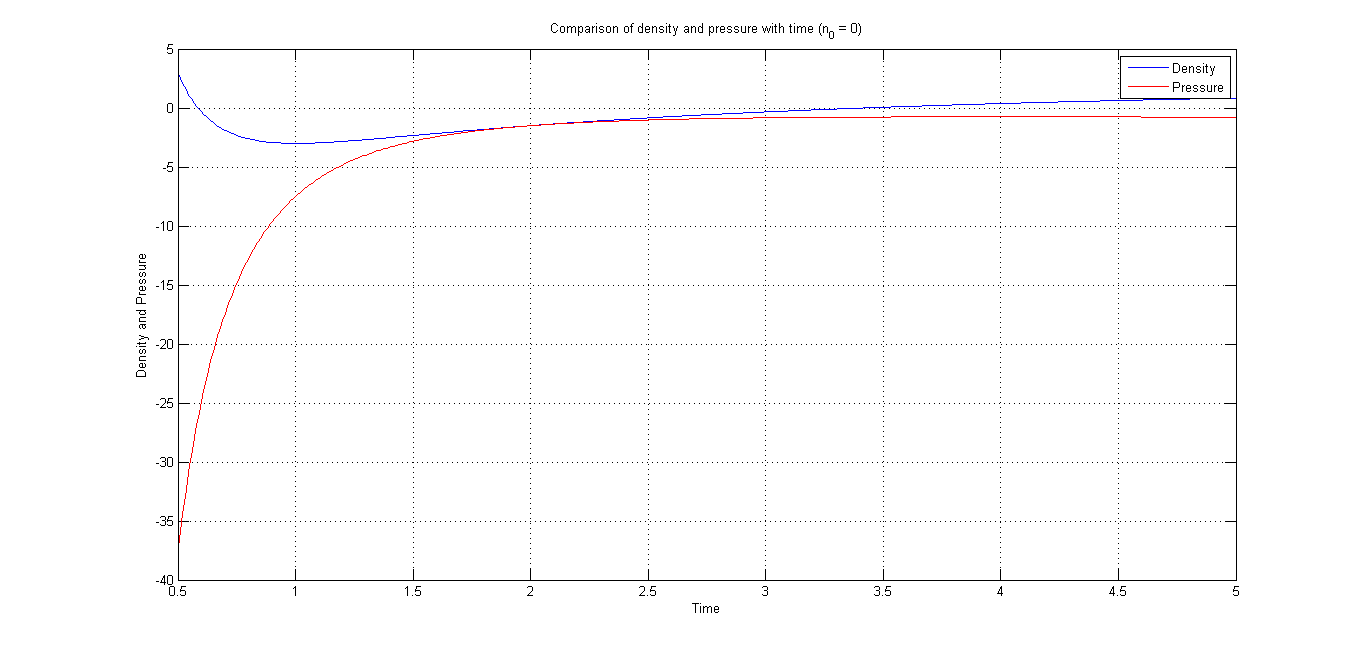}
 \caption{This figure indicates that the universe is characterized by an inflationary period in initial epoch, decelerated phase in past epoch and accelerated expansion phase in present epoch.}
\end{figure}

\end{document}